\documentclass[12pt]{ar}
\usepackage[english]{babel}
\usepackage{psfig,graphicx}
\onecolumn

\begin{document}

\title{Peculiarities and variations in the optical spectrum of the
       post-AGB star V448\,Lac = IRAS\,22223+4327}

\author{V.G.\,Klochkova, V.E.\,Panchuk, and N.S.\,Tavolganskaya}

\institute{Special Astrophysical Observatory, Russian Academy of
           Sciences, Nizhnij Arkhyz, 369167 Russia}
\date{\today}

\abstract{Repeated observations with high spectral resolution acquired in
1998--2008 are used to study the temporal behavior of the spectral line
profiles and velocity field in the atmosphere and circumstellar envelope
of the post-AGB star V448\,Lac. Asymmetry of the profiles of the strongest
absorption lines with low-level excitation potentials Ïf $\chi_{\rm
low}<1$\,eV and time variations of these profiles have been detected, most
prominently the profiles of the resonance lines of BaII, YII, LaII, SiII.
The peculiarity of these profiles can be explained using a superposition
of stellar absorption line and shell emission lines. Emission in the
(0;\,1) 5635\,\AA{} Swan band of the C$_2$ molecule has been detected in
the spectrum of V448\,Lac for the first time. The core of the H$\alpha$
line displays radial velocity variations with an amplitude of
$\Delta$Vr$\approx$8\,km/s. Radial velocity variations displayed by weak
metallic lines with lower amplitudes, $\Delta$Vr$\approx$1--2\,km/s, may be
due to atmospheric pulsations. Differential line shifts,
$\Delta$Vr\,=\,$0\div 8$\,km/s, have been detected on various dates. The
position of the molecular spectrum is stationary in time, indicating a
constant expansion velocity of the circumstellar shell, V$_{\rm
exp}$\,=\,15.2\,km/s, as derived from the C$_2$ and NaI lines.}

%\titlerunning{\it Spectroscopy of post-AGB star V448\,Lac}
%\authorrunning{\it Klochkova et al.}
\maketitle

\section{Introduction}

The semiregular variable star V448\,Lac=BD+42$^{\rm o}$4388, identifed
with the infrared source IRAS\,22223+4327 (hereafter IRAS\,22223), is one
of the most interesting and fairly well-studied protoplanetary-nebula
(PPN) candidates. Hrivnak [1] classified the star's spectral type as
Sp\,=\,G0Ia, whereas Arkhipova et al. [2] assigned a spectral type of F8I
earlier. IRAS\,22223 belongs to a small subgroup of PPN candidates with
infrared spectra containing unidentified [3] emission at
$\lambda$\,=\,21\,$\mu$~[4]. V448\,Lac displays photometric variability
typical of such objects: its maximum brightness amplitude is
0.2--0.3$^{\rm m}$ in the UBV bands [2, 5]. The spectral-energy
distribution of V448\,Lac is two-peaked, as is also typical of PPNs. The
total energy radiated in the visible is virtually the same as that
radiated in the IR [see Fig.\,5 in 4].

In the short-lived PPN evolution stage, we observe intermediate-mass stars
(with initial masses of 3--8\,M$_{\odot}$), evolving from the asymptotic
giant branch (AGB) to the planetary-nebula (PN) stage. The stellar
evolution between the AGB and PN stages from the mechanisms and
characteristic features of mass loss by AGB stars, to the complex shell
morphology of PPNs and PNs, remains incompletely understood. The role of
binarity has also been only poorly studied: it is difficult to detect a
companion of a PPN in the complex pattern of the velocity field in the
star's atmosphere and shell. The main features of the PPN evolution stage
are presented, for example, in [6]. Having consecutively passed through
evolutionary stages with core hydrogen and helium burning,
intermediate-mass AGB stars undergo a large mass loss in the form of their
strong stellar winds (the mass-loss rate can reach
10$^{-4}\,M_{\odot}/$yr). The remnant that is left after the star has lost
most of its mass has a degenerate carbon and oxygen core with a typical
mass about $0.6M_{\odot}$~[7], surrounded by an expanding gas and dust
shell, as a rule having a complex structure. Data on PPNs can be used to
investigate mass loss via stellar wind, and provide unique possibilities
for studies of stellar nucleosynthesis and the mixing and dredge-up of
nuclear-reaction products from earlier stages of the star's evolution
to the surface layers. PPN studies are also important for understanding
the evolution of galaxies, since the gas and dust ejected by such stars in
the course of their evolution enrich their galaxies in heavy elements.

Post-AGB stars evolve very rapidly. A good example is the famous star
FG\,Sge, which changed its spectral type from O3 to K2 over the past
century~[8]. Secular variations of the main parameters observed for
several PPNs have stimulated spectroscopic monitoring of the most probable
PPN candidates. Spectral variations have been detected for the optical
counterparts of IRAS\,01005+7910 [9], IRAS\,05040+4820 [10],
IRAS\,18062+2410 [11], and IRAS\,20572+4919 [12]; a trend of the effective
temperature Teff was found for the star HD\,161796\,=\,IRAS\,17436+5003
[13]. In this paper, we present the results of our highspectral-resolution
monitoring of V448\,Lac and compare these new data to those published
earlier. Our main goal is to search for possible variations of the
spectral features, and to study the velocity fields in the star's
atmosphere and shell. Section\,2 briefly describes our observation and
data-reduction techniques; Section\,3 discusses our results; Section\,4
considers stars with similar spectral characteristics; and Section 5
presents a brief summary of our main results.

\section{Observations and spectral data reduction}

Our spectroscopic data for V448\,Lac were taken at the Nasmyth focus of
the 6-m telescope BTA of the Special Astrophysical Observatory (SAO,
Russian Academy of Sciences) with the NES echelle spectrograph [14, 15].
The observations were performed using a 2048$\times$2048-pixel CCD--chip
and image slicer~[15]. The spectral resolution was R$\ge$60000 and the
signal-to-noise ratio S/N$\ge$100. The Table presents the mean observing
epochs and the measured heliocentric velocities. We extracted the
one-dimensional vectors from the 2D echelle spectra using the modified
[18] ECHELLE procedure of the MIDAS software package. We removed
cosmic-ray traces via median averaging of two spectra obtained one after
another. Our wavelength calibration was done using spectra from a Th-Ar
hollow-cathode lamp. We checked and corrected the instrumental agreement
between the spectra of the star and the hollowcathode lamp using telluric
[OI] and H$_2$O lines. The procedures used to measure the radial velocities,
Vr, from the NES spectra and the sources of errors are described in more
detail in~[19]. The rms uncertainties of the Vr measurements for stars
with narrow absorption lines are $\le$0.8 km/s (the uncertainty from a
single line).

\section{Discussion of the results}

\subsection{Peculiarity in the optical spectrum of V448\,Lac}

The main peculiarities of the optical spectrum of V448\,Lac were already
noted in the first low-spectralresolution studies. Hrivnak~[1] found that,
compared to the spectrum of a normal supergiant of a similar temperature,
the spectrum of V448\,Lac exhibited enhanced lines of s-process elements
(BaII, YII, LaII) (we consider profile peculiarities for lines of heavy
metal ions in the spectrum of V448\,Lac in more detail in Section 3.2),
and that C$_2$ and C$_3$ absorption bands were present. Our high-resolution
spectra contain the (0;\,0), (0;\,1), and (1;\,0) vibrational Swan bands of
the C$_2$ molecule. Figure\,1a shows a portion of the spectrum with the
vibrational C$_2$ (0;\,0) band of the Swan system, whose head is at
5165.2\,\AA{}. We mark there the lines of rotational transitions we used
to derive the shell expansion velocity (see Section\,3.2 below for
details). Figure\,1b demonstrates that the C$_2$~(0;\,1)~5635\,\AA{} band
in the spectrum of V448\,Lac is observed in emission. This circumstance
was not noted in~[1].

\begin{table}[t]
\bigskip
\caption{Heliocentric radial velocities, Vr, of V448\,Lac from observations in 1998--2008 Vr.
        The number of measured rotational lines of the C$_2$ Swan bands is indicated
        in parentheses. Uncertain values are printed in italics. The last line
        presents the data of [16] and [17].} 
\begin{tabular}{l|c|c|c|c|c|c}
\hline
JD=245...    &\multicolumn{6}{c}{Vr, ËÍ/Ó} \\
\cline{2-7}
            &metals & HI &\multicolumn{3}{c|}{D1,2\,NaI} & C$_2$  \\
\cline{4-6}
           & lines       &                   & 1      &   2   & 3     &    \\
\hline
1009.36    &$-$39.1&                   &$-$55.4 &$-$32.5&$-$12.4&$-57.2$(22)  \\
1273.48    &$-$42.5&$-$35.8 H$\alpha$  &$-$55.3 &\it $-$33.4&$-$12.2&         \\
2131.53    &$-$42.6&$-$40.2 H$\alpha$  &$-$55.7 &$-33.7$&$-11.8$&$-57.7$(26)  \\
3691.25    &$-$41.8&$-$40.2 H$\alpha$  &$-$55.4 &$-$29.3&$-$11.3&    \\
3692.43    &$-$41.7&$-$41.2 H$\alpha$  &$-$55.7 &$-$30.0&$-$11.5&    \\
3694.36    &$-$41.5&$-$41.0 H$\alpha$  &$-$55.3 &$-$29.2&$-$11.1&    \\
4721.15    &$-$41.8&$-32.5$ H$\alpha$  &$-$55.3 &$-27.8$&$-11.6$&    \\
4760.17    &$-$40.8&$-34.0$ H$\alpha$  &$-$55.6 &$-29.0$&$-12.2$&    \\
4774.34    &$-$41.9&$-37.8$ H$\beta$   &$-$55.8 &$-30.0$&$-11.9$&$-57.4$(84)  \\
\hline
           &$-$41.5\,[16]&\multicolumn{4}{c|}{} &$-57.2$~[17] \\ 
\hline
\end{tabular}
\end{table}

Neutral-hydrogen lines in PPN spectra are known to have peculiar profiles.
Examples of the variety of the observed profies are presented, for
example, in [20]. The complex profile of the H$\alpha$ line in the
spectrum of V448\,Lac consists of a narrow core and broad wings. Figure\,2
shows that the observed profile of this line does not agree with the
theoretical profile we computed assuming local thermodynamic equilibrium
(LTE) and using the main parameters of V448\,Lac from~[16] for a solar
hydrogen abundance. This discrepancy suggests that the shell may
contribute to the profile formation, and that the conditions under which
the H$\alpha$ core profile is formed may deviate from LTE. Figure\,3 shows
the core of the H$\alpha$ line in the spectra of V448 Lac observed on
three dates in 2005 and 2008. Figure\,3 shows that the core of the
H$\alpha$ line varies; considering the data in the Table, we conclude that
the H$\alpha$ core is systematically shifted towards shorter wavelengths
relative to the of lines of metals.

\begin{figure}[tbp]
\includegraphics[angle=-90,width=0.9\textwidth,bb=32 30 570 790,clip]{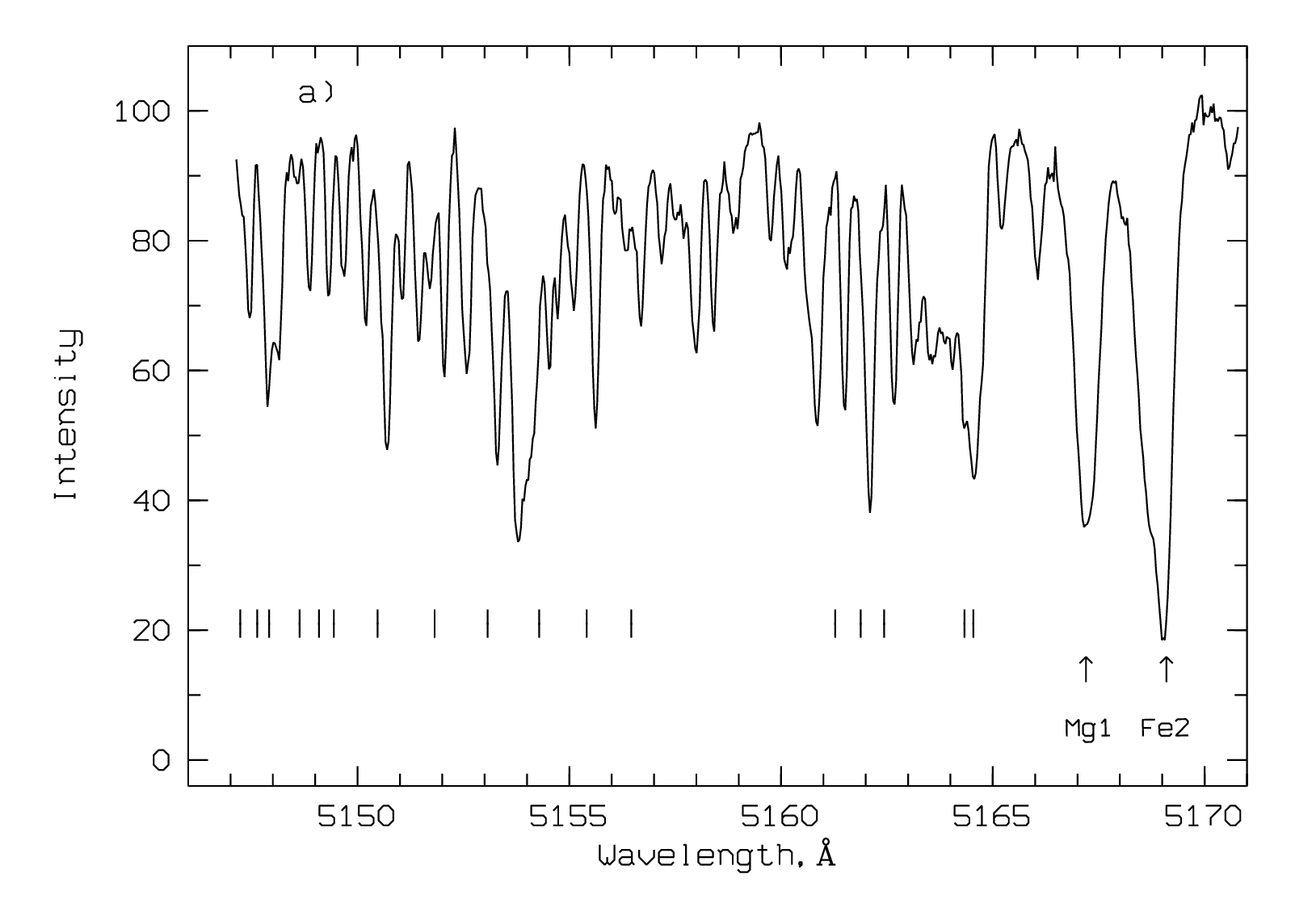}
\includegraphics[angle=-90,width=0.9\textwidth,bb=32 30 570 790,clip]{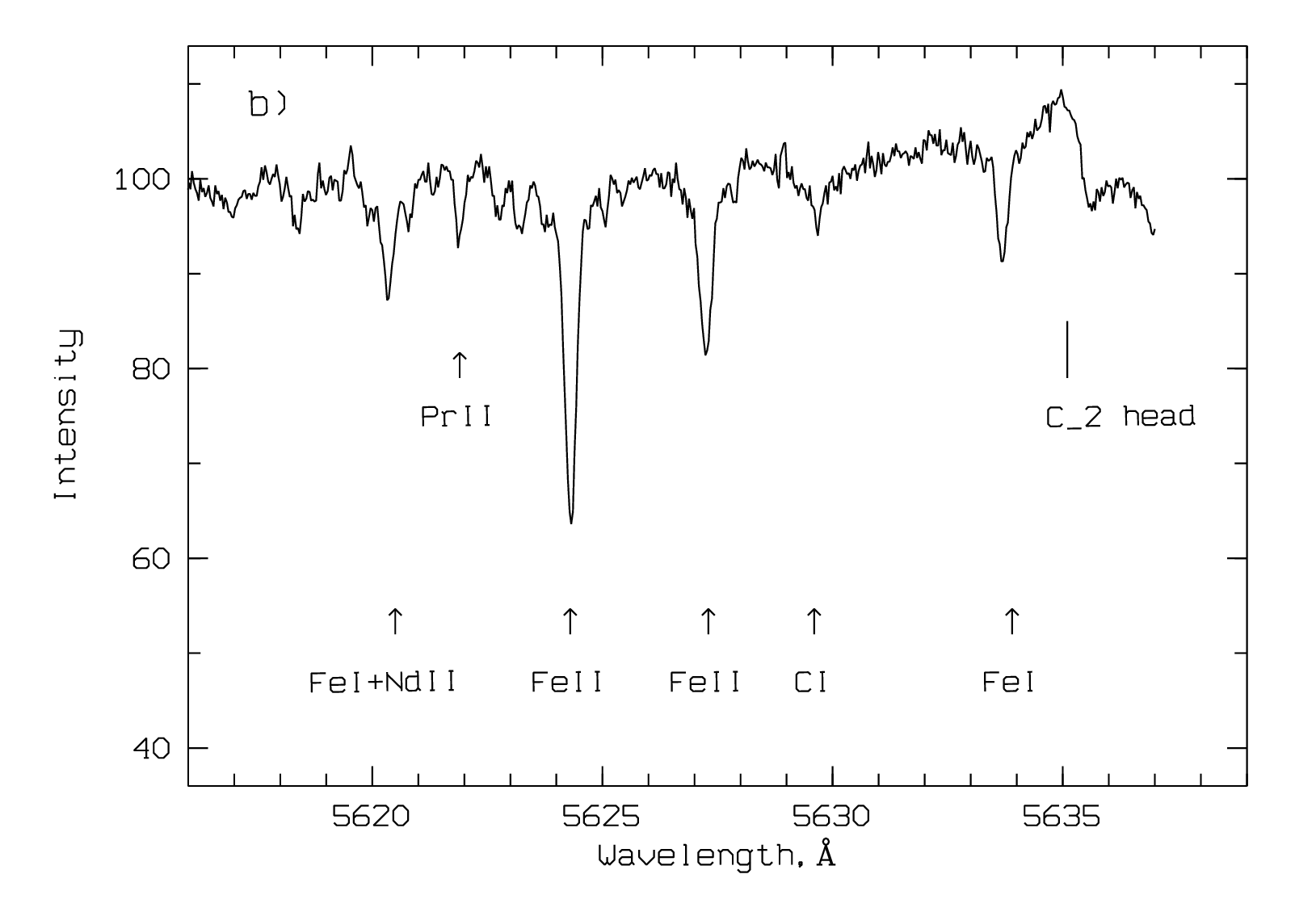}
\caption{Swan bands of the C$_2$ molecule in the spectrum of V448 Lac taken at
        JD\,2454774.34. (a) The (0;\,0) absorption band with the head at 5165 A
        (the vertical bars mark the lines of the band's rotational transitions
        used to determine the expansion rate of the shell). (b) The (0;\,1)
        emission band with the head at 5635\,\AA{}. Arrows indicate the strongest
        metal absorption lines. }
%\label{Swan}
\end{figure}

\begin{figure}[tbp]
\includegraphics[angle=0,width=0.9\textwidth,bb=32 30 570 790,clip]{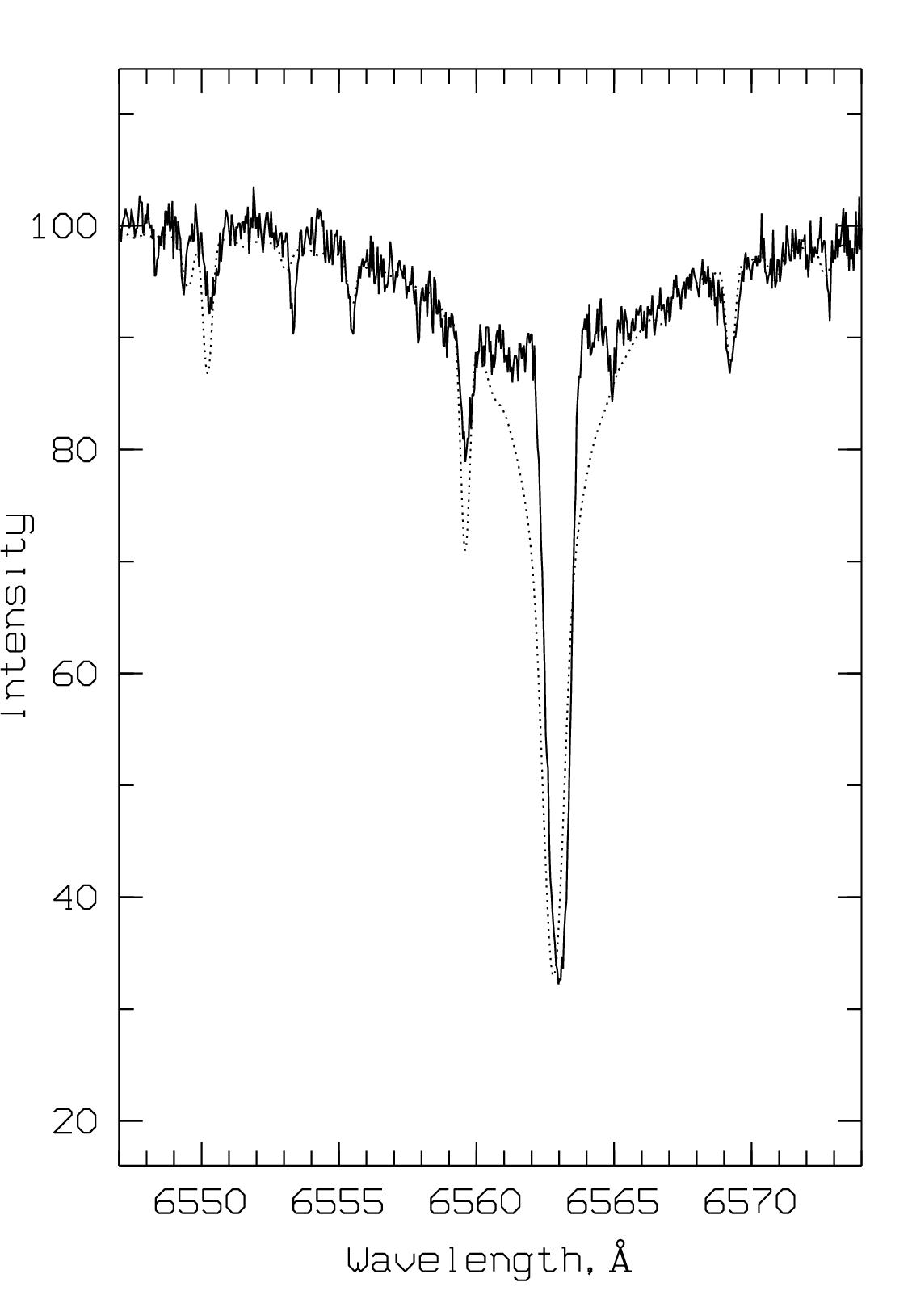}
\caption{Portion of the spectrum of V448\,Lac (JD\,2454760.17) containing
        the H$\alpha$ line (solid curve) compared to the theoretical spectrum
        (dotted curve) computed with model parameters and chemical composition
        based on data from [16].}
\end{figure}

\subsection{Asymmetric profiles of the strongest absorption lines}

The strongest absorption lines in the spectrum of V448\,Lac are
low-excitation BaII lines. It follows from Fig.\,4 that the equivalent
width W$_{\lambda}$ of the BaII\,6141\,\AA{} line is close to that of the
H$\alpha$ line. Our high spectral resolution made it possible to detect
another, previously unnoticed peculiarity of the optical spectrum of
V448\,Lac: asymmetry of the profiles of the strongest absorption lines of
ions of the heavy metals BaII, LaII, YII, and ScII, as well as of the
SiII~$\lambda$6347\,\AA{} line. This asymmetry is clearly demonstrated by
the profiles of the BaII\,6141, 6496, 5853\,\AA{} lines in Fig.\,5. All
strong BaII, LaII, YII, SiII, and ScII lines with depths
R$\rightarrow$50--70 (with the exception of hydrogen lines) display the
same kind of asymmetry: the short-wavelength wing is more extended that
the long-wavelength wing. Comparing the spectra taken on different dates,
we see that the profiles of these lines, formed in the expanding shell of
the star, vary with time as well as with line strength. The strong lines
demonstrate larger-amplitude profile variations with time, as well as
shifts of their cores towards longer wavelengths compared to the weakest
absorption lines: the velocities from the cores of these absorption lines
are within --37$<$Vr$<$--30\,km/s. Both these properties are also present
for the H$\alpha$ and H$\beta$ lines (third column of the Table):
--41$<$Vr$<$--32\,km/s.
To illustrate the profile variations of strong lines, Fig.\,6 shows the
profile of the BaII\,6141\AA{} line for 4 observing epochs.

\begin{figure}[tbp]               
\includegraphics[angle=-90,width=0.9\textwidth,bb=32 30 570 790,clip]{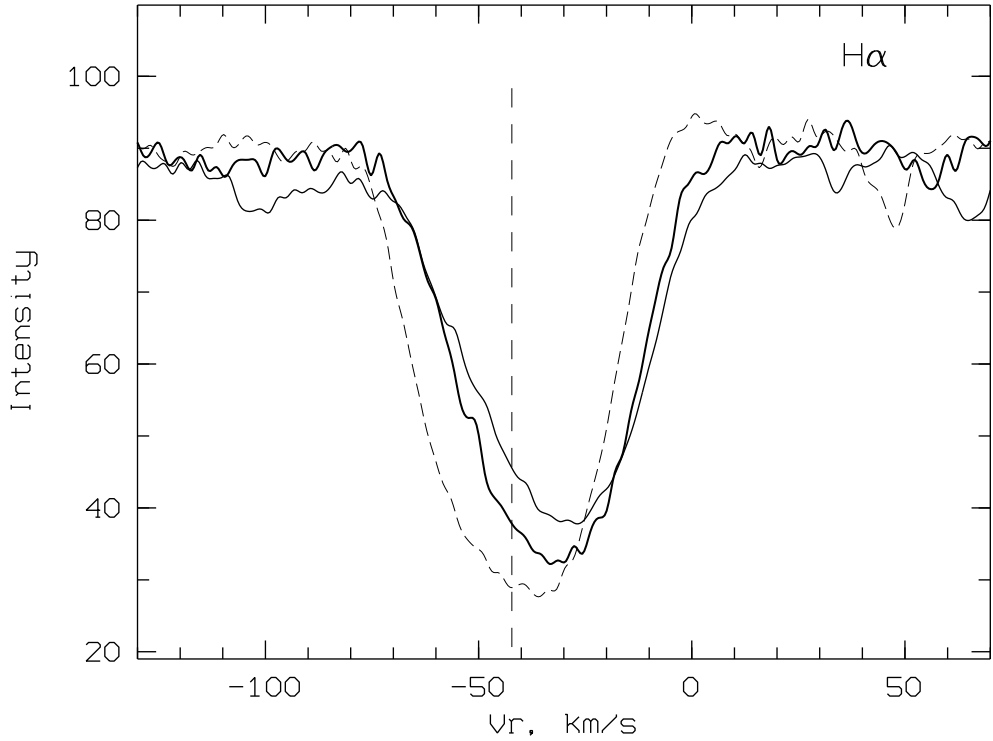}
\caption{Core of the H$\alpha$ line in the spectrum of V448\,Lac for three
         observing epochs: JD\,2454760.17 (bold), JD\,2454721.15 (thin), and
        JD\,2453694.36 (dashed). The vertical dashed line indicates the systemic
        velocity.}
\end{figure}

\begin{figure}[tbp]
\includegraphics[angle=-90,width=0.9\textwidth,bb=32 30 570 790,clip]{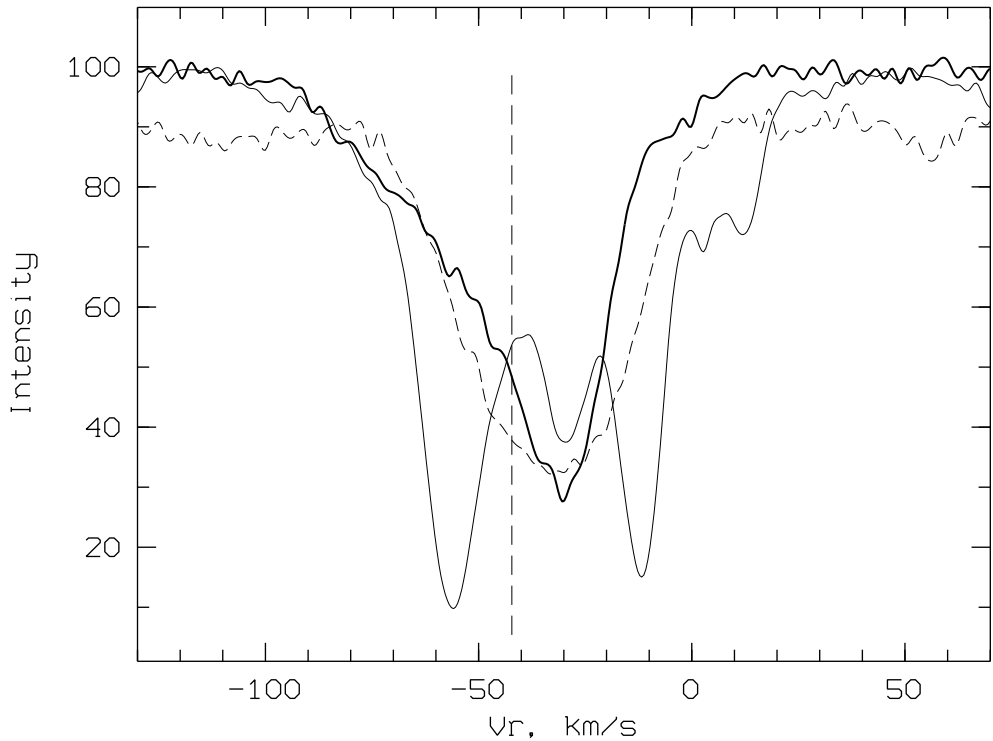}
\caption{Profiles of selected lines in the spectrum of V448\,Lac:
        BaII\,6141\,\AA{} (bold), D1\,Na (thin), and H$\alpha$ (dashed).
        The vertical dashed line indicates the systemic velocity.}
\end{figure}

Note that the variations are displayed only by the short-wavelength
profile wings and the position of the core, whereas the positions and
intensities of the long-wavelength wings of these strong BaII, LaII, and
YII absorption lines do not vary in time. An important feature is that the
radial velocity measured for the wings of these strongest absorption lines
(near the continuum) are close to the velocity derived from the weak
lines. Because of this, the radial velocity measured from the upper part
of the profile is more negative than the velocity measured for the core.
Taken together, this behavior of the profiles of the strongest lines
disagree with the profile variations expected for Schwarzschild's
mechanism~[21], which has been used to explain line splitting in the
atmospheres of some long-period variables~[22]. The stationary
longwavelength profile wings also distinguish the profile variations we
detected from those observed in the spectra of selected cepheids (see~[23]
and references therein).

In order to explain the unusual peculiarity of the profiles of strong
lines of heavy-metal ions, we substracted the asymmetric profile for the
BaII\,6141\,\AA{} line recorded on JD\,2454760.17 from the symmetric
profile observed on JD\,2453694. The results in the appearance of a
P\,Cygni-type emission line, shown in the bottom panel of Fig.\,6. The
emission line with a half-width exceeding 20\,km/s is shifted towards
shorter wavelengths relative to the systemic velocity. The emission
maximum is at Vr\,$\approx -53$\,km/s, identifying the circumstellar shell
as the probable formation region of the emission. Confirmation of this is
provided by the fact that that the epochs of the observations of V448\,Lac
when its spectrum demonstrated distorted profiles of the strong absorption
lines are also characterized by enhanced emission in the Swan C$_2$
(0;\,1) 5635\,\AA{} band. One possible mechanism for emission from the
circumstellar shell is the dissipation of energy from a shock excited by
pulsations in the stellar atmosphere [24].

\begin{figure}[tbp]
\includegraphics[angle=-90,width=0.9\textwidth,bb=32 30 570 790,clip]{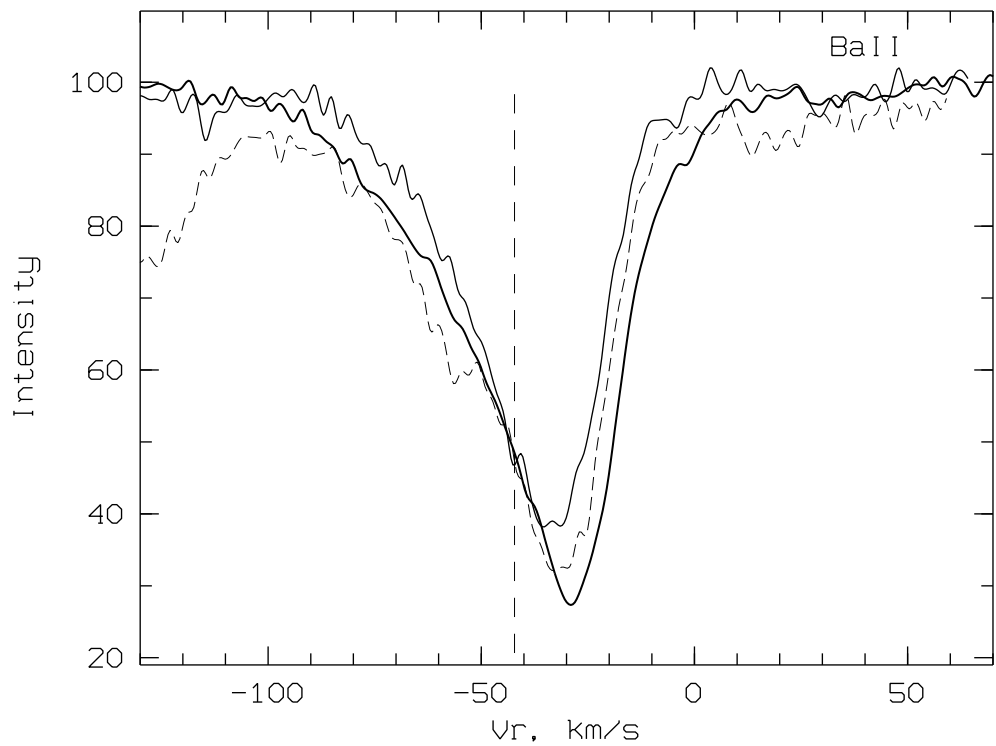}
\caption{BaII line profiles in the spectrum of V448\,Lac (observed at
        JD\,2454721.15); BaII\,5853\,\AA{} (thin), BaII\,6141\,\AA{} (bold), and
        BaII\,6496\,\AA{} (dashed). The vertical dashed line indicates the systemic
        velocity.}
\end{figure}

Similar emission lines of the ScII, TiII, FeII, YII, and BaII ions with
widths of approximately 10--20\,km/s and shifted towards shorter
wavelengths are known for pulsating R\,CrB stars near their brightness
minima [25--27]. In the spectrum of R\,CrB, such emission features appear
in the cores of absorption lines after a decrease of the star's brightness
by 3$^{\rm m}$ [26]. It is believed that the emission is formed in the
upper atmospheres (chromospheres?) and shells of R\,CrB stars, due to the
passage of shocks generated by pulsations. However, the atmospheric
pulsations in the case of V448 Lac\,have low amplitudes, and it remains
unclear how shocks could form, since this requires a much higher pulsation
amplitude, $\Delta$Vr$\ge$5--10\,km/s [28].

The profile variations we found in the spectrum of V448\,Lac could be due
to a non-spherical shape and/or non-uniform structure of its circumstellar
shell. At epochs when the observer receives radiation that is not
distorted by the shell, the star's spectrum displays symmetric absorption
profiles. At epochs when the light from the star comes through and is
scattered by the circumstellar matter of the expanding gas and dust shell,
the shell's emission profile is superposed on the profiles of the
strongest absorption lines. In fact, the IR image of the nebula around
V448\,Lac obtained with a spatial resolution of about 1$^{''}$ is
aspherical. Meixner et al.~[29] suggested it was a rare type of toroidal
shell. Observations with the Hubble Space Telescope in the visible with a
higher resolution, 0.27$^{''}/$pixel [30], detected more detailed
structure of the nebula, in the form of 2 pairs of lobes.

\begin{figure}[tbp]
\includegraphics[angle=-90,width=0.9\textwidth,bb=32 30 570 790,clip]{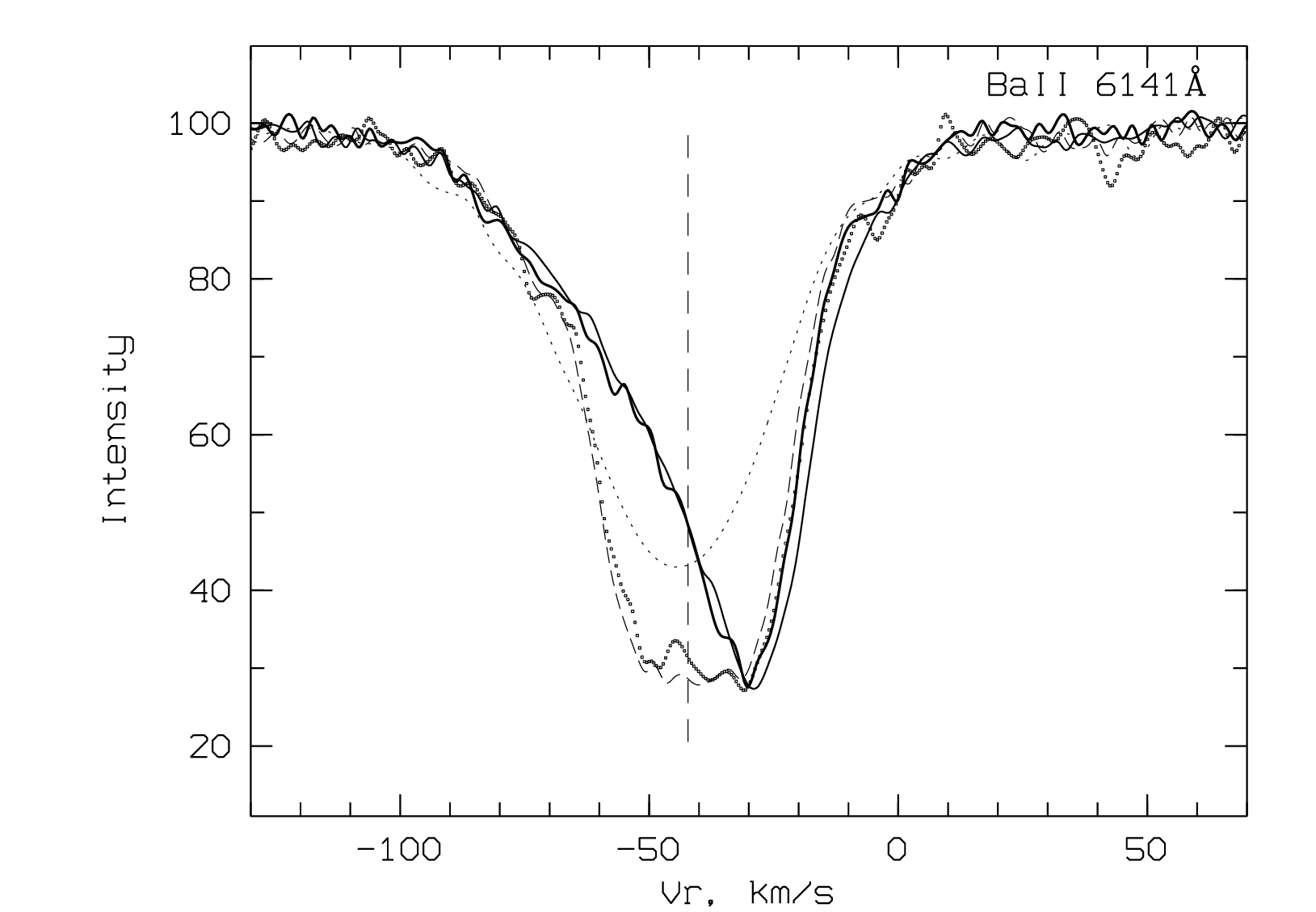}
\includegraphics[angle=-90,width=0.9\textwidth,bb=32 30 570 790,clip]{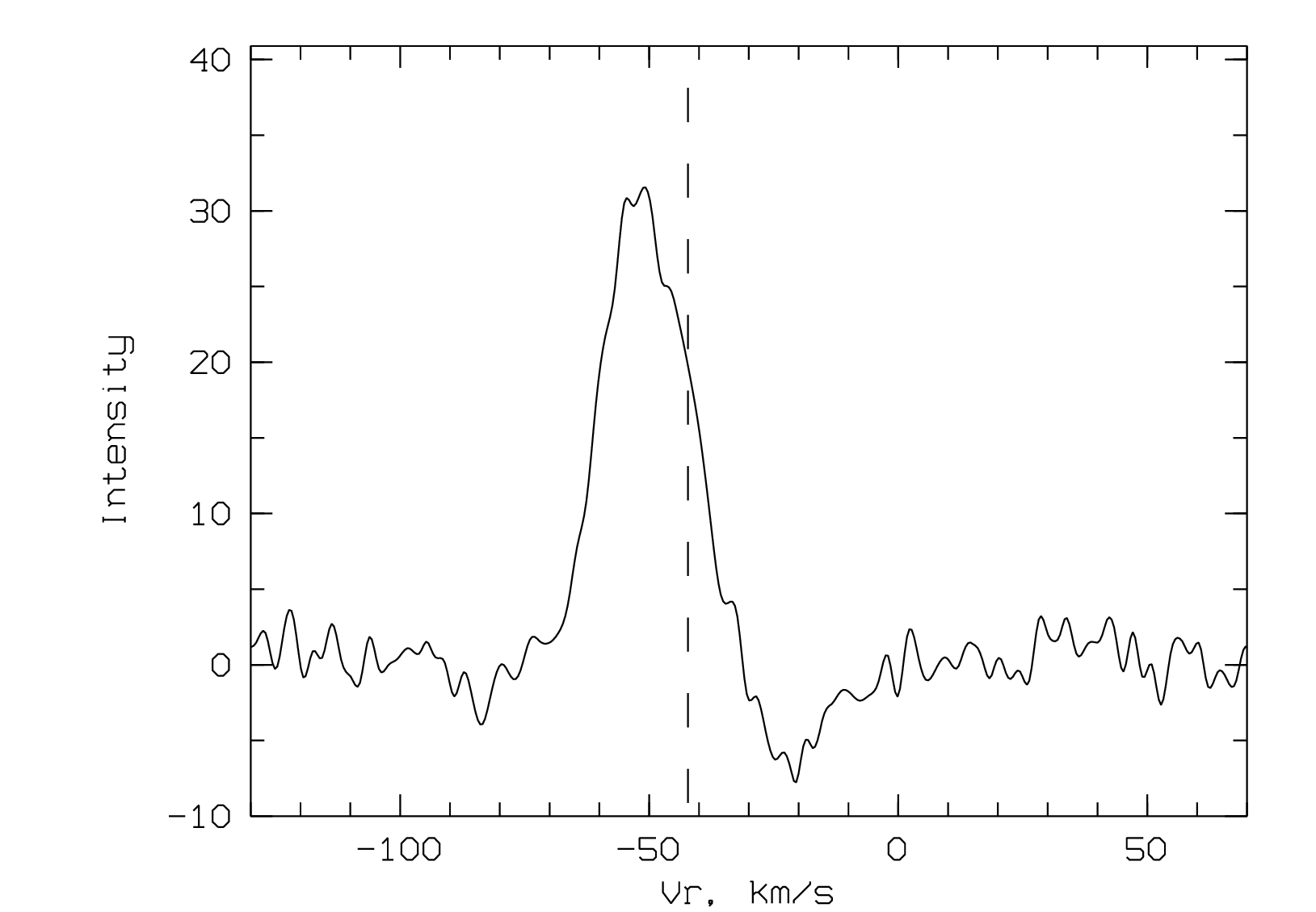}
\caption{Top: profile variations of the BaII\,6141\,\AA{} line in the
         spectra of V448\,Lac at various epochs: JD\,2454760.17 (bold),
         JD\,2454721.15 (thin), JD\,2453694.36 (dashed), and JD\,2452131.53 (dotted).
         Bottom: emission component of the BaII\,6141\,\AA{} line profile in the
         spectrum of V448\,Lac. The vertical dashed line indicates the systemic
        velocity.}
\end{figure}

\subsection{Radial velocity variations lines of metals}

To derive the mean radial velocity, Vr, we measured a large number of the
least blended absorption lines in the spectra of V448\,Lac (300--400 lines
in each spectrum). We selected the lines using a spectral atlas for the
post-AGB star HD\,56126, which can be considered a canonical post-AGB
object [31]. The atlas was prepared by Klochkova et al.~[32] using echelle
spectra taken with the same NES spectrograph on the 6-m SAO telescope. The
amount of blending in the spectrum of V448\,Lac is higher than in the
spectrum of HD\,56126 for two reasons: first, the effective temperature is
lower for V448\,Lac (Teff\,=\,6500\,K [16]) than for HD\,56126
(Teff\,=\,7000\,K [33]), and second, the metallicity of V448\,Lac ([Fe/H]$
= -0.4$\,dex [16]) is higher than that of HD\,56126 ([Fe/H]$=-1.0$\,dex
[33]).

Because of the stronger blending, the accuracy of Vr measurements from
single lines (the rms deviation, $\sigma$) is somewhat worse for our
V448\,Lac spectra than for stars with narrow, medium-intensity absorption
lines: about 1.5\,km/s. In the Table, the data from the 2005--2008 spectra
are complemented with our Vr measurements from spectra obtained in
1998--2001 at the 6-m telescope using the ``Lynx'' echelle spectrograph
[34] with a resolution of R\,=\,25000. The last line of the Table also
presents the mean Vr values from [16]. This measurement is important for
us because the spectrum was acquired earlier than our first observations.
It follows from the long-term observations of Hrivnak~[5] that the
amplitude and period of the radial-velocity variations of V448\,Lac are
typical of PPNs: the radial velocity varies from --36 to --43\,km/s with
the period of 89$^{\rm d}$. All our mean Vr values for metallic absorption
lines in the Table are in a narrower range, from --39 to --42\,km/s.

Taking into account the large probability of differential motions in the
outer atmospheric layers of V448\,Lac, we tabulated Vr values for
individual lines and groups of lines. The second column of the Table
presents the mean velocity derived from absorption lines of low and
intermediate intensity, Vr(Met). This restriction of the line sample is
related to the revealed differential line shifts (see Section\,3.4 for
more detail). Subsequent columns of the Table contain the velocities
derived from the H$\alpha$ or H$\beta$ lines, the short-wavelength shell
component (column ``1''), the atmospheric component (column ``2''), the
longwavelength interstellar component (column ``3'') of the D lines of the
NaI doublet (cf. the observed profile of the NaI doublet lines in
Fig.\,7), and rotational lines of the C$_2$ Swan bands. Let us consider
these results in more detail.

{\it Neutral hydrogen lines}. The velocity measured from the cores of the
H$\alpha$ and H$\beta$ differs systematically from the mean for lines of
metals, namely they are shifted towards shorter wavelengths. Note also the
larger variation amplitude of the radial velocity derived from the
H$\alpha$ core ($\Delta$Vr$\,\approx$\,8\,km/s) compared to the variations
of the mean velocity from weak metallic lines
($\Delta$Vr$\,\approx$\,1--2\,km/s).

{\it Molecular Spectrum.} Besides the IR--excess and reddening, the
presence of a gas and dust shell around the central stars of PPNs is
revealed by molecular features in the optical spectra of these objects.
Since molecular spectra can be formed in stellar atmospheres with
temperatures Teff\,$\le$\,3000 K, they should be formed in circumstellar
shells surrounding G0 supergiants. In our recorded wavelength range, the
spectra of V448\,Lac exhibit vibrational Swan bands of C$_2$. Our high
spectral resolution enables accurate measurements of the positions of the
rotational lines of the Swan bands. Using the rotationalline wavelengths
from the electronic tables of Bakker et al.~[17], we measured the
positions of several dozen rotational lines of the (0;\,0) and (1;\,0) Swan
bands and determined the mean radial velocity in the band formation
region. Since their profies are narrow compared to those of atmospheric
lines, the rotational lines of the (0;\,0) Swan band are easy to identify
in the spectrum of V448\,Lac (Fig.\,1a). Thus, the uncertainty in the
positions of single lines ($\le$\,0.8\,km/s) is much lower than that for
atmospheric absorption lines. Since head (1;\,0) Swan band in the
short-wavelength part of the spectrum, 4712--4734\,AA{}, is more strongly
blended with atmospheric lines, the measurement accuracy is somewhat
worse: $\sigma \approx$\,0.9\,km/s. Our velocity measurements from
rotational components of the vibrational Swan bands of C$_2$ gave the mean
value Vr(C$_2$)\,=\,--57.4\,km/s ($\sigma \le$\,0.8\,km/s) from the
rotational lines of the C$_2$~(1;\,0) band with its head at 4734\,\AA{}
and from the lines of the C$_2$~(0;\,0) band with its head at 5165\,\AA{}.
The mean velocity for our earliest observation (at JD\,2451009.36) is
Vr(C$_2$)\,=\,--57.2\,km/s. These measurements are in excellent agreement
with the data of Bakker et al.~[17]: Vr(C$_2$)\,=\,--57.2\,km/s.

The shift of the circumstellar features relative to the systemic velocity
enables us to determine the expansion rate of the corresponding regions in
the circumstellar shell relative to the systemic velocity, V$_{\rm lsr}$.
The systemic velocity of IRAS\,22223, V$_{\rm lsr}=-30.0$\,km/s (the
heliocentric systemic velocity is V$_{sys} = -42.2$\,km/s) was determined
as the velocity of the center of the CO(2--1) emission profile at
millimeter wavelengths [35]. In contrast to the CO emission lines formed
in the extended envelope expanding in all directions, the observed
molecularcarbon absorption lines are formed in a restricted part of the
shell located between the star and observer. As a result, we obtain the
expansion velocity of the shell (i.e., of the part where the Swan bands
are formed), V$_{\rm exp}$\,=\,15.2\,km/s. This can be interpreted as the
expansion velocity derived from the optical spectra; it agrees well with
the expansion rate for IRAS\,22223, V$_{\rm exp}=15.0$\,km/s, from the
catalog~[36], where numerous observations of circumstellar shells in the
CO and HCN molecular bands are collected. Note also that the
shell-expansion rate of IRAS\,22223 is typical of the circumstellar shells
of post-AGB stars (see, for instance,~[36]).

In the case of V448\,Lac, the heliocentric velocity from the metal lines
is close to the systemic velocity, V$_{\rm sys} =-42.2$km/s. This provides
evidence for the absence of a stellar-mass secondary in the IRAS\,22223
system. This result is non-trivial because the chemical evolution,
mixing, and dredge-up of the accumulated products of nuclear reactions to
the surface atmospheric layers, matter outflow, and the formation of
the shell morphology could be different is the presence of a secondary.

{\it NaI\,D lines.} The lines of the resonance NaI doublet in the spectrum
of V448\,Lac have a complex structure. It follows from the Table and
Fig.\,4, which shows the profile of the D1 line, that the doublet lines
contain three strong absorption components located at the velocities
Vr(NaI)\,=\,--55.8, --30.0, and --11.9\,km/s. The component with
Vr\,=\,--55.8\,km/s is probably formed in the circumstellar shell, where
the shell C$_2$ Swan bands are formed. The component with
Vr\,=\,--11.9\,km/s (V$_{\rm lsr}=--27$\,km/s) is interstellar. The
presence of this interstellar component in the spectrum of V448\,Lac
suggests that the star is located behind the local spiral arm in the
Galaxy. It follows from [37] that the radial velocity in the local arm is
V$_{\rm lsr} \approx -10$\,km/s, while in the Perseus arm, it is V$_{\rm
lsr} \approx -55$\,km/s. Thus, the distance to the Perseus arm can be
taken as an upper estimate for the distance to V448\,Lac. Taking into
account Galactic CO maps~[38] and the Galactic coordinates
(l\,=\,96.75$^{\rm o}$, b\,=\,$-11.5^{\rm o}$) and systemic velocity
(V$_{\rm lsr} = -30.0$\,km/s) of IRAS\,22223, we obtain for the estimated
distance of the source from the Sun $d \approx$\,3.5 kpc.

It is more difficult to determine where component ``2'' in the system,
whose velocity is Vr$\approx -30.0$\,km/s, is formed. To determine this,
we compared the observed profile of the NaI~D lines to the theoretical
profile computed using model parameters corresponding to the results
from~[16]. For our spectral synthesis, we applied the SynthV code
developed by Shulyak et~al.~[39] and adapted by its authors for use in a
Linux operating system. We computed a flat-parallel model with the
stellar parameters from~[16] using the code of~[39]. Comparing the
profiles in Fig.\,7, where the position of the theoretical profile
corresponds to the mean velocity from atmospheric absorption lines of
metals, we can see that component ``2'' resembles the atmospheric NaI line
with its core distorted similarly to the cores of strong BaII lines. This
is confirmed by the similarity of the BaII\,6141\,\AA{} line profile and the
profile of the NaD component with Vr\,=\,--30\,km/s in Fig.\,4.

\begin{figure}[tbp]
\includegraphics[angle=0,width=0.9\textwidth,bb=32 30 570 790,clip]{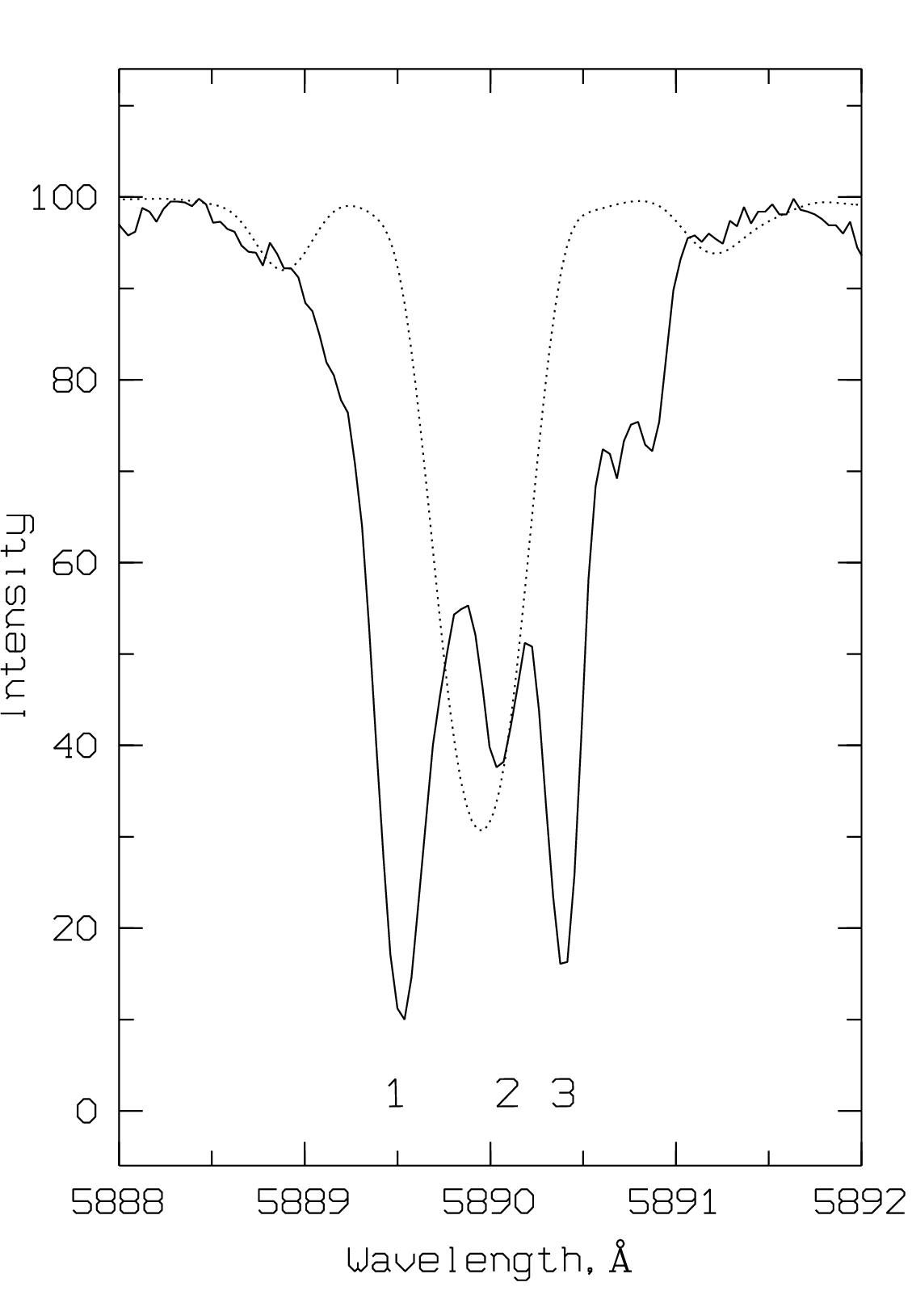}
\caption{Comparison between the observed profile of the NaI doublet in
     the spectrum for JD\,2454734.34 (solid curve) and a theoretical profile
     computed with the model parameters and chemical composition corresponding
     to the data from [16] (dotted curve). The components of the observed
     profile are marked: ``1'' is the shell component, ``2'' the atmospheric
     component, and ``3'' the interstellar component.}
\end{figure}

\subsection{Differential line shifts and probable pulsations in
           the atmosphere}

In the PPN case, in addition to Vr variations with time, the
radial-velocity changes can also be complicated by differential motions in
the extended atmospheres of these objects. A detailed analysis of the
radial velocities from spectra with high spectral and temporal resolution
for selected brightest PPNs enables the detection of behavior differences
for radial velocities determined from lines with different excitations,
formed at different depths in the stellar atmosphere. The extended
spectral range recorded using our echelle spectrograph enabled us to study
the differential radial velocities measured for lines of different
intensities for V448\,Lac. We plotted Vr(R) diagrams for each of our
spectra of V448\,Lac -- relations between the heliocentric radial
velocity, Vr , measured for the core of an absorption line and its central
depth, R. Note that the continuum level corresponds to zero depth, while
the strongest lines have R\,$\rightarrow$\,100. Some examples of such
relations typical of V448\,Lac are shown in Fig.\,8. The star's
center-of-mass velocity, V$_{\rm sys} = -42.2$\,km/s~[35], derived from CO
observations in the millimeter, is indicated by the horizontal dashed
lines in the panels. In all the diagrams in Fig.\,8, the circumstellar and
atmospheric components of the NaI doublet are marked, as well as the cores
of the H$\alpha$ and H$\beta$ lines and of resonance lines with abnormal
profiles. Panel (d) also indicates the velocity from the rotational
lines of the CO Swan system for the mean depth of these lines.

Klochkova and Chentsov~[40] earlier analyzed the radial velocities of
HD~56126 using a large set of lines with various natures (weak atmospheric
absorption lines, HI lines, molecular bands, profile components of the
NaI\,D lines) in spectra taken on different dates. These data revealed
radial-velocity variations of HD\,56126 even for the weakest absorption
lines [40, Fig.\,7], confirming the pulsational nature of the
radial-velocity variations. By analogy with HD\,56126, this suggests that
the small radial-velocity variations of V448\,Lac are due to low-amplitude
pulsations. In this case, we expect variability of Vr from all lines
formed in the star's atmosphere, including low-intensity lines.
Actually, Fig.\,8 confirms the presence of Vr variations for weak lines
in the spectrum of V448\,Lac: the mean radial velocity derived for the
spectrum taken on JD\,2454774.34 from weak lines (R\,$\le$\,20) coincides
with the systemic velocity of V448\,Lac, while the weak-line radial
velocity for a spectrum obtained two weeks earlier (JD\,2454760.17) is
systematically below the systemic velocity. The difference is not large
($\Delta Vr \approx  1$--2\,km/s), but is significant, given the small
uncertainties of our velocity measurements. Thus, variations of the
star's radial velocity were also detected for the weakest absorption
lines (with depths R\,$\le$\,20) in the spectra of V448\,Lac, a possible
manifestation of low-amplitude pulsations. The presence of pulsations
should be no surprise for V448\,Lac, which has an extended atmosphere and
is located in the instability strip. Hrivnak and Lu~[5] also concluded
earlier that the Vr and brightness variations of V448\,Lac had a
pulsational character. In addition to the fact that the brightness and
radial velocity vary, these authors noted the irregular character of the
variations. Note that, in contrast to our data, which is differentiated
according to line groups, the radial velocities of~[5] were derived using
a spectrometer, without taking into account line intensity differences.

Figure 8 shows that the dynamic state of the atmosphere was different at
the different observing epochs considered: three of the panels (b, c, d)
show special positions of the cores of the neutral hydrogen and
low-excitation BaII, LaII, and YII lines, whose profile peculiarity was
already mentioned in Section\,3.1. However, the positions of all the
stellar lines in the spectrum for JD\,24543964.36 (panel a), independent
of their depth, corresponds to the systemic velocity of the star, while
the observations for JD\,2454760.17 reveal differential line shifts up to
$\Delta$\,Vr\,$\approx8$\,km/s that depend on the line intensities.

\section{Spectra of related objects}

Features originating in circumstellar shells have long been known in the
spectra of high-mass F and G hypergiants with IR--excesses (see, for
example, [41, 42]). It follows from~[41] that the yellow hypergiant
$\rho$\,Cas, whose luminosity is close to the Humphreys--Davidson
limit, is characterized by a complex, timevariable radial-velocity pattern
due to the radial pulsations. One of the results of~[41] is the special
temporal behavior of the low-excitation absorption lines, in
particular, the resonance BaII\,6141\,\AA{} line, whose core contains a
circumstellar component. The high-excitation lines behave regularly with
time, whereas the BaII lines show deviations from such behavior. In a
restricted phase range, the spectrum of $\rho$\,Cas exhibits split cores of
the low-excitation lines, probably due to the presence of a
variable-intensity emission component. A detailed study of the spectral
variations of $\rho$\,Cas and a comparison with long-period R\,CrB variables
and post-AGB supergiants was recently presented by Gorlova et al.~[43].
These authors note that current models of atmosphere dynamics cannot
explain the absorption-core splitting and the appearance of emission lines
in the spectra of cool pulsating supergiants~[43]. In addition to taking
into account dynamic phenomena, it is necessary to consider non--LTE
excitation mechanisms for spectral lines.

In the case of low-mass supergiants in the post--AGB stage, the influence
of the circumstellar shell on the profiles of lines with low excitation
potentials for their lower levels was long known only for the D--lines of
the resonance doublet of sodium, NaI. Parthasarathy et al. [44] were the
first to note the abnormal profile of the BaII\,6141\,\AA{} line in the
spectrum of the post--AGB star HD\,56126. We recently published
spectroscopy of the post-AGB star V354\,Lac where we have detected
manifestations of the star's shell in the profiles of the heavy-metal ions
BaII, LaII, CeII, and NdII for the first time [45]. Let us consider the
properties of these two stars compared to V448\,Lac in more detail.

\subsection{Analogy to V354\,Lac}

The semiregular variables V448\,Lac and V354\,Lac (identified with the
IR--source IRAS\,22272+5435) have similar main parameters, spectral types
and luminosity classes, and large IR--excesses. The IR--spectra of both
stars display emission at $\lambda$\,=\,21\,$\mu$. A comparison of the
chemical abundances determined for V448\,Lac in~[16] and for V354\,Lac in
our recent paper~[45] also reveals similar peculiarities: their
atmospheres are depleted in iron, and show large excesses of carbon and
heavy metals. The combination of these properties makes it certain that
both stars are in the transition to planetary nebulae. The IR observations
of~[46] indicate that the morphology of the nebulae around V354\,Lac and
V448\,Lac are toroidal. The systemic velocities of the two stars are also
similar, and the same component of interstellar sodium, at Vr\,=\,$-12\div
-13$\,km/s, is identified in both their spectra.
In addition, like that of V448\,Lac, the spectrum of V354\,Lac displays
abnormal profies for lines of heavy-metal ions with low-level excitation
potentials $\chi_{\rm low} < 1$\,eV~[45]. The anomaly of the profiles in
the spectrum of V354\,Lac is more strongly expressed, and is sometimes
observed as split absorption cores. The only significant difference
between the spectra of V448\,Lac and V354\,Lac is that all the Swan bands
in the spectrum of V354 Lac\,are in absorption, while we see the
C$_2$\,(0;\,1) 5635\,\AA{} band in emission in the spectrum of V448\,Lac.

\subsection{Analogy to HD\,56126}

The characteristic peculiarities observed for V448\,Lac (its brightness
variations, instability of the radial velocities of atmospheric lines, and
the spectral peculiarities in the IR and optical, including molecular
features and the variable and peculiar H$\alpha$ profile), as well as the
similarity of the elemental abundances in the atmospheres of the two
stars, suggests a similarity between this star and the post-AGB star
HD\,56126\,=\,IRAS\,07134+1005. HD\,56126 has been thoroughly studied due to its
high visual brightness.
Like V448\,Lac, HD\,56126 is a member of a small subgroup of PPNs that
exhibit emission at 21\,$\mu$. The members of this PPN subgroup have
central-star atmospheres enriched in carbon and sprocess heavy metals (cf.
[16, 20] for details). It is interesting that the shell morphology is also
toroidal for this object [47]. In general, HD\,56126 can be considered a
canonical post-AGB object.

Like for V448 Lac, the H$\alpha$ line in the spectrum of HD\,56126 has an
abnormal profile, whose longterm variations were studied by Oudmaijer
and Bakker [48]. Barthes et al. [49] conducted a detailed study of
variations of the brightness and atmospheric velocity field of HD\,56126.
Having analyzed extensive photometric and spectroscopic monitoring data
for HD\,56126, Barthes et al. [49] determined a radial pulsation period of
P\,=\,36.8$^{\rm d}$, with the brightness and radial-velocity
semi-amplitudes being 0.02\,$^{\rm m}$ and 2.7\,km/s. They emphasize that
the passage of pulsation-driven shocks leads to complex dynamics in the
upper atmospheric layers where the cores of hydrogen lines are formed.
Thus, the peculiarity of provide crucial information that can help reveal
the structure of the stellar environs.

In addition to the similar properties of 448\,Lac and HD\,56126, we also
note certain differences between the spectroscopic features of interest.
One of the main results of our spectroscopic study of V448\,Lac is the
discovery of asymmetry and variability of the profiles of strong
atmospheric absorption lines of BaII etc. However, the photospheric
components of the Na\,D double in the spectrum of HD\,56126 do not reveal
such peculiarities~[49]. Nevertheless, noting the asymmetry of the BaII
profiles in the spectrum of HD\,56126, Parthasarathy et al.~[44] suggested
that there is a contribution from shell component.

\section{Conclusions}

We have detected asymmetry of strong absorption lines with lower-level
excitation potentials $\chi_{\rm low} < $1 eV in optical spectra of the
post-AGB star V448\,Lac obtained using the echelle spectrograph of the
6-meter SAO telescope with a spectral resolution of R\,$\ge$\,60 000. This
is most notably true of the resonance absorption lines of the BaII, YII,
LaII ions. We detected profile variabilities of these lines with time.
These profile peculiarities can be explained as superpositions of spectral
features: absorption lines formed in the star's atmosphere and shell
emission lines.

The observed profile of the H$\alpha$ line, whose core varies with time,
disagrees with the theoretical profile computed for the star's main
parameters and a normal (solar) hydrogen abundance. This suggests the
presence of a contribution by a shell to the profile and/or deviations of
the formation conditions for the H$\alpha$ profile from LTE. We found
radial-velocity variations in the H$\alpha$ core with amplitude
$\Delta$\,Vr$\approx 8$\,km/s. Variations of the mean velocity with a lower
amplitude, $\Delta$\,Vr$\approx 1-2$\,km/s, were also detected for the
weakest absorption lines in the spectra of V448\,Lac (with depths of
R\,$\rightarrow $\,20), a possible manifestation of low-amplitude
pulsations. At some epochs, we found differential shifts in the positions
of lines with different intensities.

The position of the molecular spectrum does not change with time,
indicating a stationary shell expansion rate. The expansion velocity of
the shell in the formation region of the C$_2$ Swan bands is V$_{\rm
exp}$\,=\,15.2\,km/s.

Finally, we emphasize that explaining the variations of Vr and the line
profiles in the spectra of V354\,Lac and V448\,Lac and the development of
models for these systems require further studies of these (and related)
objects, including spectroscopic monitoring and spectropolarimetry.
Spectropolarimetry with high spectral resolution can provide crucial
information that can help reveal the structure of the stellar environment.

\section*{Acknowledgments}

This study was supported by the Russian Foundation for Basic Research
(project code 08--02--00072\,a) and the Basic Research Program of the
Presidium of the Russian Academy of Sciences ``The Origin, Structure,
and Evolution of Objects in the Universe''.

\begin{figure}[tbp]
\includegraphics[angle=-90,width=0.6\textwidth,bb=32 30 570 790,clip]{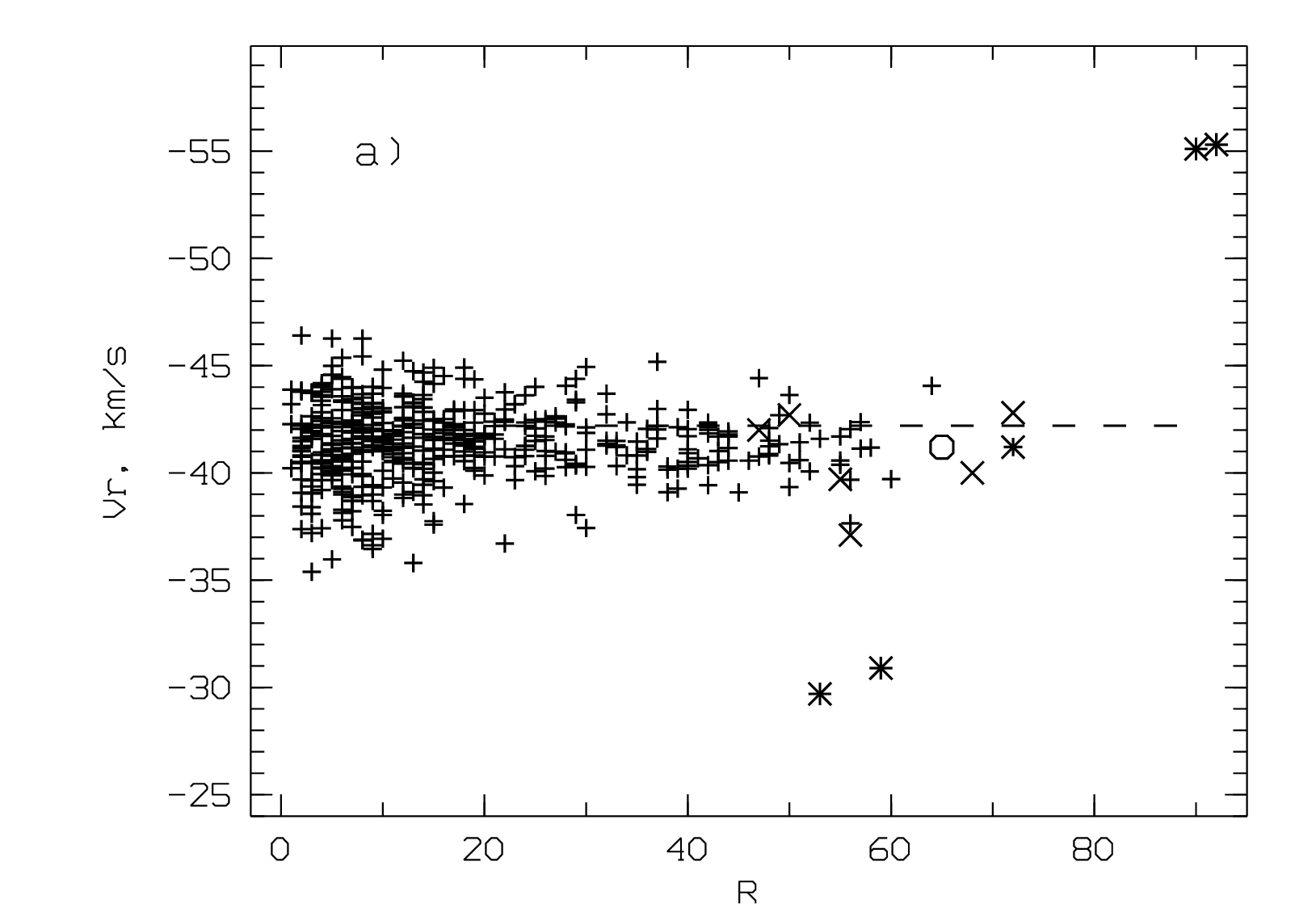}
\includegraphics[angle=-90,width=0.6\textwidth,bb=32 30 570 790,clip]{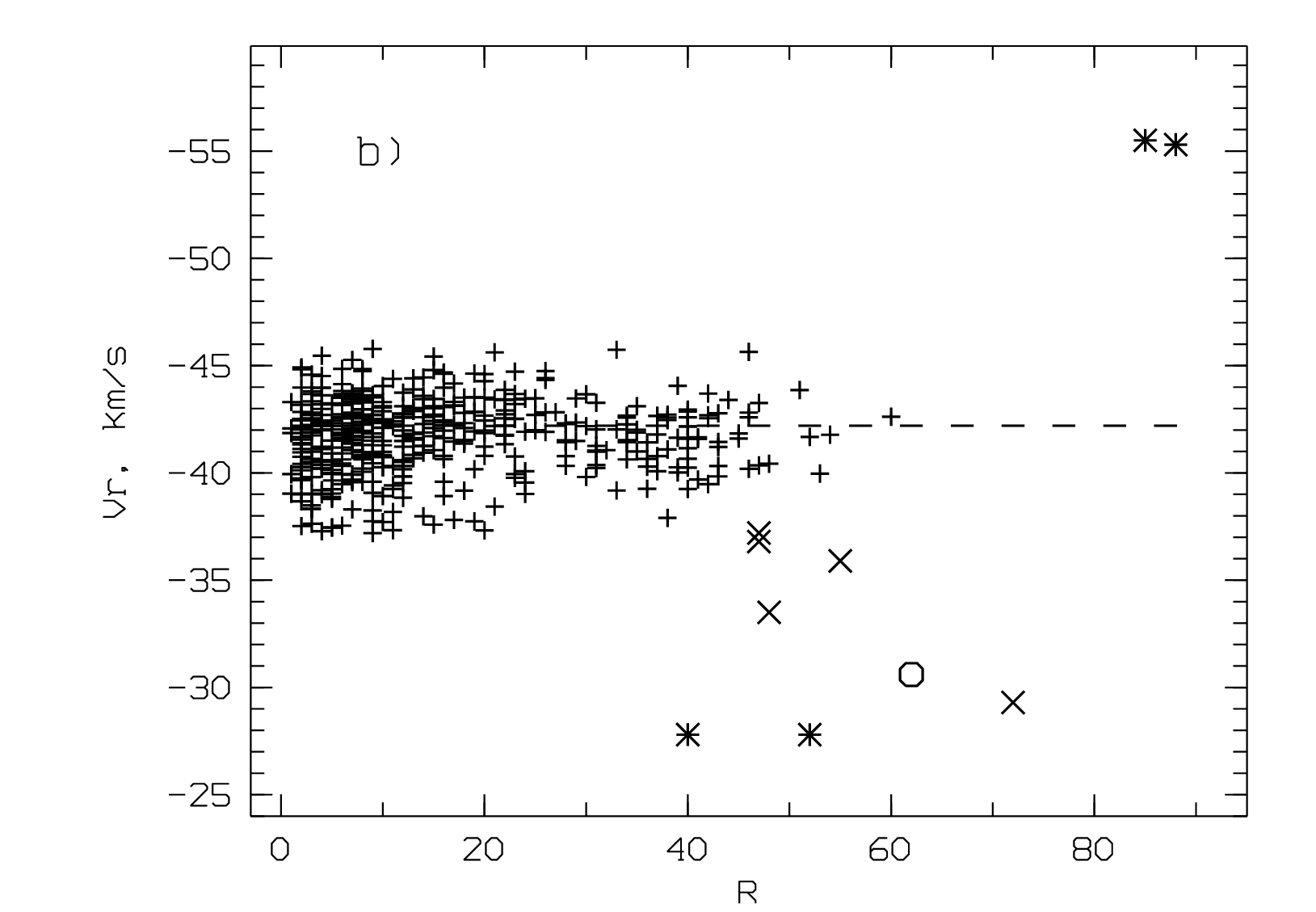}
\includegraphics[angle=-90,width=0.6\textwidth,bb=32 30 570 790,clip]{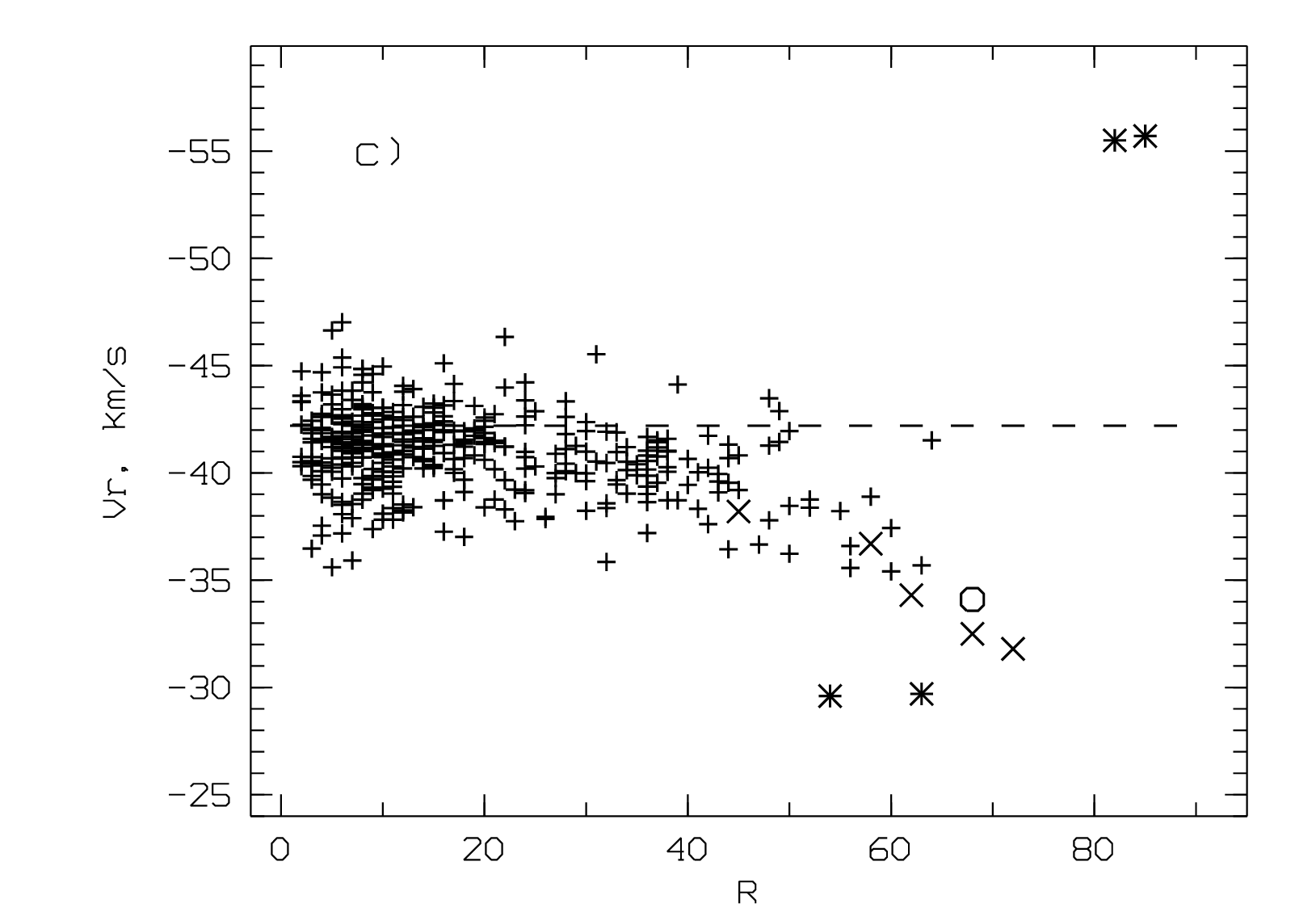}
\includegraphics[angle=-90,width=0.6\textwidth,bb=32 30 570 790,clip]{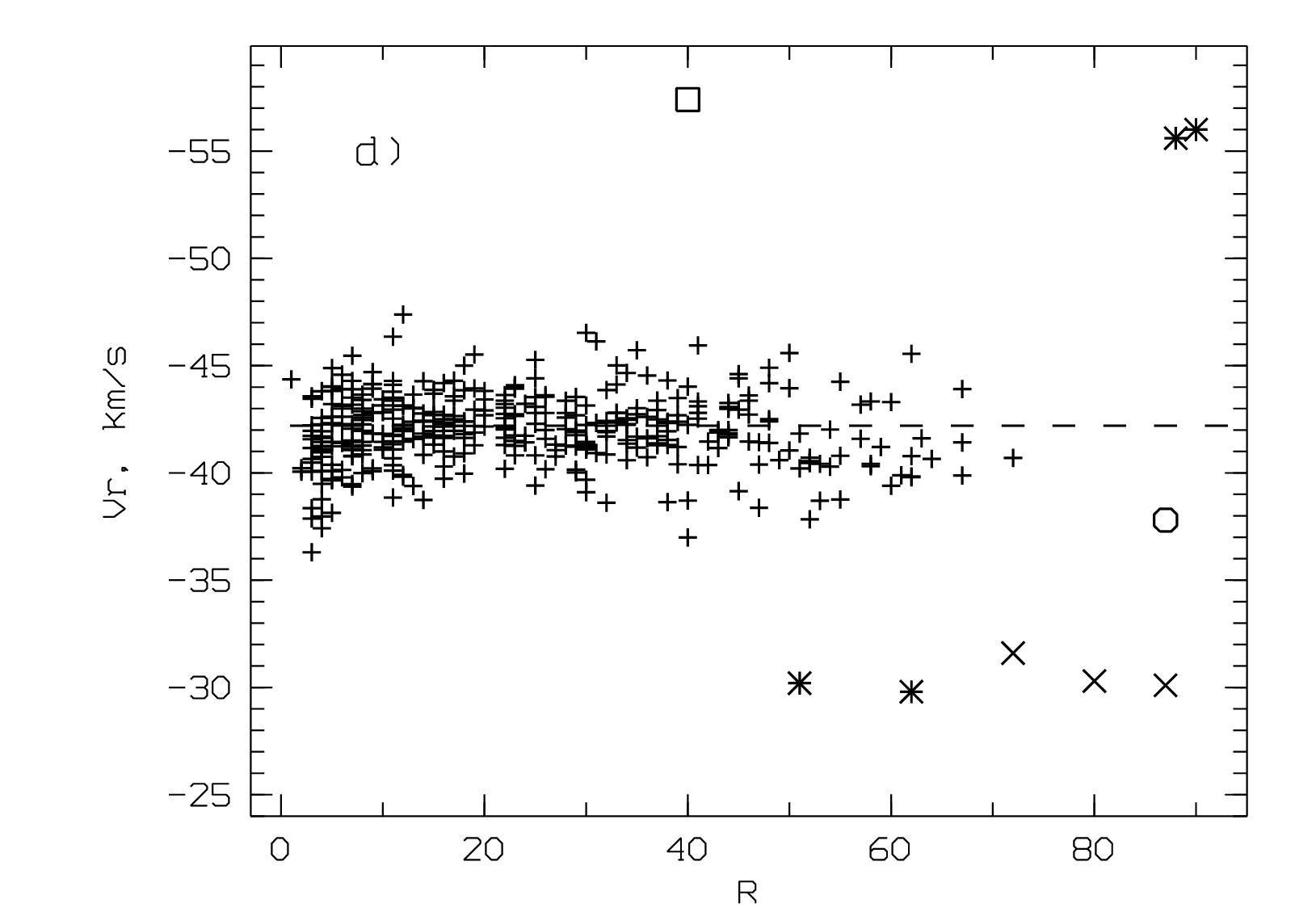}
\caption{Relation between the heliocentric radial velocity determined for
   the absorption core and the absorption's depth, R, in the spectra of
   V448\,Lac. The dashed line indicates the systemic velocity. The pluses show
   lines of metals, with each symbol representing one line. The asterisks are
   the interstellar (in the upper parts of the panels) and atmospheric (in
   the lower parts of the panels) components of the NaI doublet.
   (a) JD\,2453964.36; crosses are the cores of the BaII\,5863, 6141, 6496\,\AA{},
   ZrII\,5349\,\AA{}, YII\,5402\,\AA{}, ScII\,5526\,\AA{}, and
   SiII\,6346\,\AA{} lines, the circle is the H$\alpha$ line.
   (b) JD\,2454721.15; crosses are the cores of the BaII\,5863, 6141\,\AA{},
   ZrII\,5349\,\AA{}, YII\,5402\,\AA{}, and ScII\,5526\,\AA{} lines,
   the circle is the H$\alpha$ line.
   (c) JD\,2454760.17; crosses are the cores of the BaII\,5853, 6141,
   6496\,\AA{}, LaII\,6389\,\AA{}, and YII\,5402\,\AA{} lines, the
   circle is the H$\alpha$ line. (d) JD\,2454774.34; crosses are the cores
   of the BaII\,4554, 4934\,\AA{} and YII\,5123\,\AA{} lines, the circle is the
   H$\beta$ line, and the square is the mean for the rotational lines of the C$_2$
   Swan bands. The sizes of the symbols correspond to the accuracy of the
   measured velocities and line depths.}
\end{figure}

\end{document}